\documentclass[amsmath,trackchanges,twocolumn]{aastex702}

\usepackage{physics}

\begin{document}

\title{A Multi-instrument Generalized Lomb-Scargle Periodogram Implementation}

\author[orcid=0000-0001-8171-1127]{Giulia Martos}
\affiliation{Max-Planck-Institut für Astronomie, Königstuhl 17, 69117 Heidelberg, Germany}
\affiliation{Fakultät für Physik und Astronomie, Universität Heidelberg, Im Neuenheimer Feld 226, 69120 Heidelberg, Germany}
\affiliation{Department of Astronomy, Universidade de São Paulo, Rua do Matão 1226, 05508-090 São Paulo, Brazil}
\email[show]{gimartos@mpia.de}  

\author[orcid=0000-0001-9513-1449]{Néstor Espinoza} 
\affiliation{Space Telescope Science Institute, 3700 San Martin Drive, Baltimore, MD 21218, USA}
\affiliation{William H. Miller III Department of Physics and Astronomy, Johns Hopkins University, Baltimore, MD 21218, USA}
\email{nespinoza@stsci.edu}

\author[orcid=0000-0002-4933-2239]{Jorge Meléndez} 
\affiliation{Department of Astronomy, Universidade de São Paulo, Rua do Matão 1226, 05508-090 São Paulo, Brazil}
\email{jorge.melendez@iag.usp.br}

\begin{abstract}
We present an implementation of the well-known and broadly used Generalized Lomb-Scargle Periodogram (GLS) which simultaneously takes into account individual offsets introduced in the data by different instruments: a ``multi-instrument" GLS periodogram. While the algorithm itself is not new, the simple closed form implementation we provide, to our knowledge, is. We showcase an application of our multi-instrument GLS on radial-velocity data of the hot Jupiter TOI-481~b, whose period is recovered with a higher significance compared to the standard GLS approach. The code is available on GitHub. 

\end{abstract}

\section{Introduction} 

    Searching for periodic signals in astronomical data is one of the most common problems in time-series analyses. Usages range from searches on photometric data to study a wide variety of phenomena, including, e.g., stellar activity or pulsation, to radial-velocity (RV) searches of exoplanetary reflex-motion signals. The key framework used to search for those is that of Lomb-Scargle (LS), which introduces a technique that iterates over a wide range of trial periods on a time-series dataset, returning the amplitude of sinusoids at each (a so-called ``periodogram") to aid in such searches \citep{Lomb_1976, Scargle_1982}. Implementations of this periodogram framework vary widely \citep[see, e.g.,][]{Ferraz-Mello_1981} but undoublty one of the most popular in the past 20 years is the Generalized Lomb-Scargle (GLS) periodogram of  \cite{Zechmeister_Kuerster_2009}, which accounts for a floating offset on a dataset simultaneously with the amplitude of the trial period. 
    
    In RV exoplanet searches, measurements obtained with different instruments and at different epochs are often combined to increase the observational baseline and enhance the sensitivity and precision of exoplanet detection studies. These typically have their own RV instrumental offsets. The original GLS periodogram, however, assumes a single offset for the dataset at hand. A common approach to try to overcome this is to subtract the median or mean RV from each instrument's dataset, but the validity of this approach depends strongly on the data quality and time baseline. For instance, a mean-substracted dataset that measured RVs at a particular set of points in phase-space might remove useful RV information when combined with other datasets that measured the RV curve in a longer portion in phase-space. This might, thus, impact on the detectability of certain trial periods when performing GLS. An alternative approach is to, at each trial period, fit for not only the amplitudes and a single offset, but for \textit{all} offsets of different instruments simultaneously to avoid this loss of information.

    Here we present an implementation of the GLS periodogram, the \textit{multi-instrument} GLS periodogram, which takes into account individual offsets introduced by the different instruments in the data set simultaneously for a given trial period. We note that our model is not necessarily new. The periodogram introduced in \cite{baluev:2008} already accounts for a general linear model that can be added at each trial period; an extra offset per instrument is just a special case of this model. In a similar way, our implementation is also a special case of the multi-band algorithm of \cite{vi:2015}. This latter one fits a much more complex fourier series to data, which needs a regularization parameter to control the fit. In our case, we strictly follow the original, simpler GLS methodology and simply add a single mean to each instrument --- this makes our algorithm simple, closed form and in principle as speedy and numerically stable as the original GLS; the code of this implementation is available on \href{https://github.com/Giumartos/Multiple-instrument-periodogram}{Github}. Here, we showcase its applicability on the RV dataset of TOI-481~b, a dataset on which our multi-instrument GLS periodogram retrieves the correct period, whereas the original GLS implementation is unable to.  

\section{Method}

    We adopted the least-squares interpretation of the periodogram \citep{VaderPlas_2018}, in which the data are modeled using sinusoidal functions evaluated over a range of frequencies, and the periodogram measures the goodness of fit of the model at each frequency. The highest peak therefore corresponds to the frequency that provides the best explanation of the periodic signal present in the time series. Including a constant offset in the sinusoidal model provides additional flexibility and allows the periodogram to more reliably identify the correct period. This is particularly relevant for RV data obtained with different instruments, for which the constant offset represents the instrument-dependent zero-point. Consequently, this offset cannot be assumed to be the same for the entire dataset, as is commonly done in standard periodogram analyses.

    To derive the equations for the multi-instrument periodogram, we followed a procedure similar to that presented by \cite{Zechmeister_Kuerster_2009}. The main difference is that, rather than assuming a common offset for the sinusoidal functions across all instruments, we introduced an independent offset for each instrument. Further details of the derivation are provided in the Appendix (Section \ref{sec:appendix}).

\section{Application to the system TOI-481}

    To test the method, we used the tool to search for periodic signals in the time series of the TOI-481 system, which is known to host a hot Jupiter with an orbital period of $\sim 10.33$ days \citep{Brahm_2020_TOI481}. We used the RV data made available in the discovery paper, obtained with six different instruments. For comparison, we also computed a standard GLS periodogram using the \texttt{LombScargle} function from \texttt{astropy.timeseries} \citep{astropy_2013, astropy_2018, astropy_2022}, using the same frequency grid as that adopted for the multi-instrument periodogram. We computed the standard GLS periodogram both using the uncorrected RVs and after subtracting the median RV of each instrument.

\begin{figure*}[ht]
    \plotone{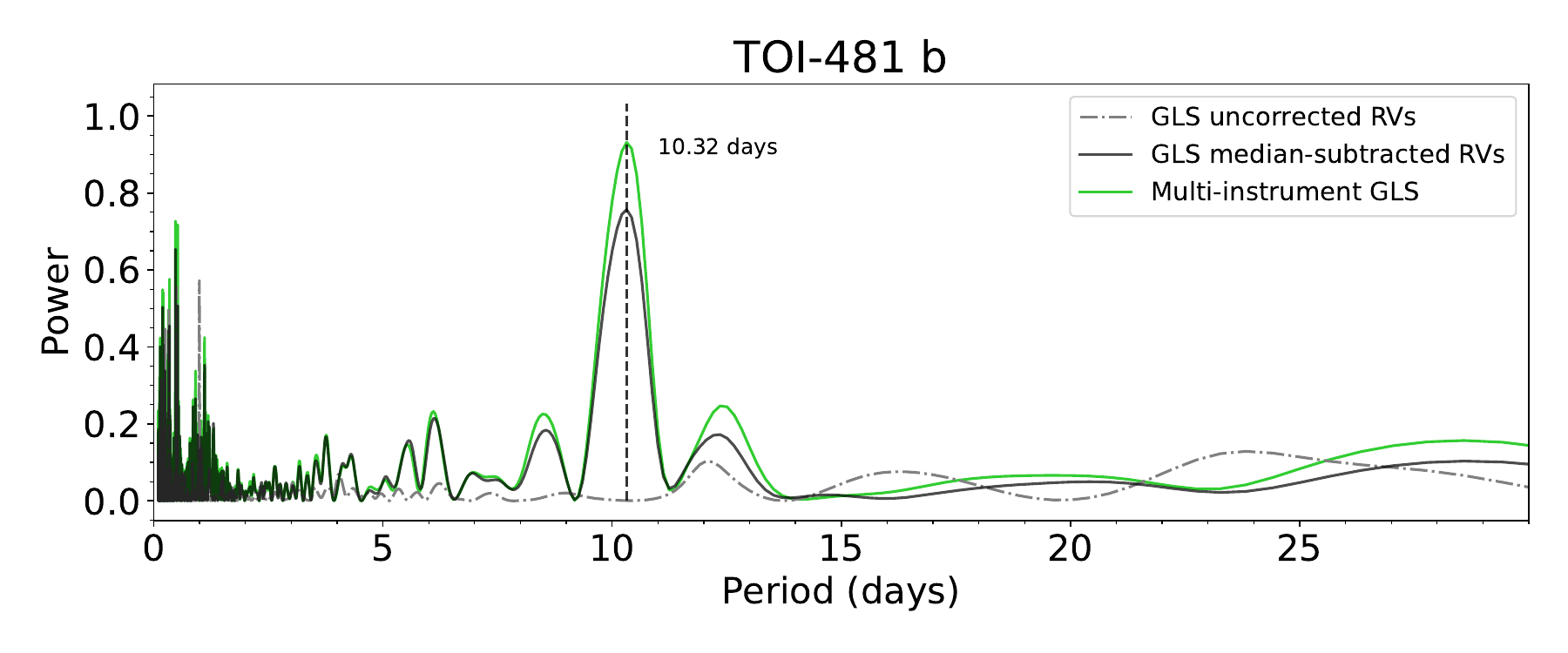}
    \caption{Comparison of the standard Lomb-Scargle periodogram (GLS) computed using the uncorrected and instrument-by-instrument median-subtracted data with the multi-instrument periodogram.}
    \label{fig:comparison}
    \end{figure*}

    As shown in Figure \ref{fig:comparison}, the multi-instrument GLS periodogram successfully recovered the planet's orbital period, with a higher significance than that obtained with the standard GLS periodogram applied to the median-subtracted data.

\section{Code availability}

    The code is available on \href{https://github.com/Giumartos/Multiple-instrument-periodogram}{GitHub} and Zenodo \citep{giulia_martos_2026_21842269} with an example dataset.

\begin{acknowledgments}
    G.M. and J.M. acknowledge FAPESP (São Paulo State Research Foundation) for the grants 2019/22664-4 and 2019/09249-8.
\end{acknowledgments}

\clearpage

\appendix 
    \section{Derivation of the periodogram}
    \label{sec:appendix}

    The following notation is used to derive the power function of the multi-instrument periodogram. Here, $y_{k,i}$ denotes the $i$-th measurement obtained with instrument $k$ at time $t_{k,i}$, with an associated uncertainty $\sigma_{k,i}$. $N_k$ is the number of data points obtained with instrument $k$, and $\omega$ is the angular frequency, defined as $\omega = 2\pi f$, where $f$ is the frequency:

	\begin{eqnarray}
	W_{k} = \frac{1}{{\sigma_{k,i}}^2} \label{notW} \\ 
    w_{k,i} = \frac{1}{{W_{k} \sigma_{k,i}}^2} \label{notw} \\
	\hat{C_{k}} = \sum^{N_{k}}_{i=1}w_{k,i} cos(\omega  t_{k,i}) \label{notC} \\ 
    \hat{S_{k}} = \sum^{N_{k}}_{i=1}w_{k,i} sin(\omega  t_{k,i}) \label{notS} \\
    \hat{Y_{k}} = \sum^{N_{k}}_{i=1}w_{k,i}  y_{k,i} \label{notY} \\
    \hat{CC} = \sum^{M}_{k=1}W_{k} \sum^{N_{k}}_{i=1} w_{k,i} {cos}^2(\omega t_{k,i}) \label{notCC}\\
    \hat{SS} = \sum^{M}_{k=1}W_{k} \sum^{N_{k}}_{i=1} w_{k,i}  {sin}^2(\omega t_{k,i}) \label{notSS} \\ 
    \hat{CS} = \sum^{M}_{k=1}W_{k} \sum^{N_{k}}_{i=1} w_{k,i} cos(\omega t_{k,i}) sin(\omega  t_{k,i}) \label{notCS} \\ 
    \hat{YS} = \sum^{M}_{k=1}W_{k} \sum^{N_{k}}_{i=1} w_{k,i}  y_{k,i} sin(\omega  t_{k,i}) \label{notYS}\\ 
    \hat{YC} = \sum^{M}_{k=1}W_{k} \sum^{N_{k}}_{i=1} w_{k,i} y_{k,i} cos(\omega t_{k,i}) \label{notYC}
	\end{eqnarray}		

    The sinusoidal model, $y = a\cos(\omega t) + b\sin(\omega t) + c$, is fit over a range of parameter combinations by minimizing the $\chi^2$ for the $M$ instruments (Equation \ref{eq:chi2}). The total $\chi^2$ is defined as the sum of the individual $\chi^2_k$ values for each instrument (Equation \ref{eq:chi2_k}).

	\begin{eqnarray}
	\chi^2_{k} = W_k \sum^{N_{k}}_{i=1}w_{k,i}[y_{k,i} - y_{k}(t_{k,i})]^2 \label{eq:chi2_k} \\ 
    \chi^2 = \sum^{M}_{k=1} \chi^2_{k} = \sum^{M}_{k=1}W_k \sum^{N_{k}}_{i=1}w_{k,i}[y_{k,i} - y_{k}(t_{k,i})]^2 \label{eq:chi2} \\ 
    y_{k} = acos(\omega t_{k,i}) + bsin(\omega t_{k,i}) + c_{k} \label{eq:y_k} 
	\end{eqnarray}	
	
	To minimize Equation \ref{eq:chi2}, we calculate the partial derivatives with respect to the coefficients $a$ and $b$ and the instrument-dependent offsets $c_k$, and set them equal to zero, resulting in the following equations:

	\begin{itemize}
	\item Derivative with respect to $a$
	\begin{eqnarray}
    a \hat{CC} + b\hat{CS} + \sum^{M}_{k=1} c_{k}  W_{k}  \hat{C_{k}} = \hat{YC}
	\end{eqnarray}

	\item Derivative with respect to $b$
	
	\begin{eqnarray}
	a \hat{CS} + b\hat{SS} + \sum^{M}_{k=1} c_{k} W_{k} \hat{S_{k}} = \hat{YS}
	\end{eqnarray}

	\item Derivative with respect to $c_{k}$

	\begin{eqnarray}
	a \hat{C_{k}} + b\hat{S_{k}} + c_{k} = \hat{Y_{k}}
	\end{eqnarray}

    \end{itemize}
	
    The coefficients $a$ and $b$, together with the $M$ instrument-dependent offsets $c_k$, are obtained by solving the resulting linear system of $M+2$ equations (Equation \ref{eq:system}).
	
	\begin{eqnarray}
	\begin{bmatrix}
    \label{eq:system}
	\hat{CC} & \hat{CS} & W_{1} \hat{C_{1}} & W_{2}  \hat{C_{2}} & ... & W_{M} \hat{C_{M}}\\ \hat{CS} & \hat{SS} & W_{1} \hat{S_{1}} & W_{2}\hat{S_{2}} & ... & W_{M} \hat{S_{M}}\\ \hat{C_{1}} & \hat{S_{1}} & 1 & 0 & ...& 0\\ \hat{C_{2}} & \hat{S_{2}} & 0 & 1 & ... & 0 \\ ... & ... & ... & ... & ... & ...\\ \hat{C_{M}} & \hat{S_{M}} & 0 & 0 & ...& 1 
	\end{bmatrix}
	\cdot
	\begin{bmatrix}
	a\\b\\c_{1}\\c_{2}\\...\\c_{M}
	\end{bmatrix} 
	=
	\begin{bmatrix}
	\hat{YC}\\ \hat{YS}\\ \hat{Y_{1}}\\ \hat{Y_{2}}\\...\\ \hat{Y_{M}}
	\end{bmatrix} 
	\end{eqnarray}\\
	
	We then calculate the periodogram power of each frequency over the range of frequencies of interest using Equation \ref{eq:power}.
	
	\begin{eqnarray}
	p(\omega) = \frac {\chi^2_{0} - \chi^2(\omega)}{\chi^2_{0}}
	\label{eq:power}
	\end{eqnarray}	 
	
	\begin{eqnarray}
	\chi^2_{0} = \sum^{M}_{k=1}W_{k} \sum^{N_{k}}_{k=1} w_{k,i}[y_{k,i} - c_{k}]^2
	\label{chi2_0}
	\end{eqnarray}

\bibliography{references.bib}{}
\bibliographystyle{aasjournalv7.1}

\end{document}